\documentclass[fleqn]{annalen}

\usepackage{graphicx}

\begin{document}

\newcommand{\volume}{8}              %sets current volume,
\newcommand{\xyear}{1999}            %sets year in header
\newcommand{\issue}{5}               %sets current issue,
\newcommand{\recdate}{29 July 1999}  %sets received date,
\newcommand{\revdate}{dd.mm.yyyy}    %sets revised date,     
\newcommand{\revnum}{0}              %number of revisions,
\newcommand{\accdate}{dd.mm.yyyy}    %sets accepted date,
\newcommand{\coeditor}{ue}           %sets (co)editor,
\newcommand{\firstpage}{507}         %first page number,  
\newcommand{\lastpage}{510}          %last page number,
\setcounter{page}{\firstpage}        %sets page counter to first page number 

%%%%%%%%%%%%%%%%%% please give up to three keywords here %%%%%%%%%%%%%%%%%%%%%

\newcommand{\keywords}{disordered solids, quantum wires, conductance
distribution} 

%%%%%%%%%%%%%%%% please give up to three PACS numbers here %%%%%%%%%%%%%%%%%%%

\newcommand{\PACS}{71.25.-s, 72.15.Rn, 05.45+b}

%% please enter (First) Author (et al.) and short version of the title here %%
%%%%%%%%%%%% must not exceed 80 characters in length together %%%%%%%%%%%%%%%%

\newcommand{\shorttitle}{P. W\"olfle et al. Conductance
distribution of disordered quasi one-dimensional wires} %% sets the header on oddpage

%%%%%%%%%%%%%%%%%%%%%%%% here comes the title group %%%%%%%%%%%%%%%%%%%%%%%%%%

\title{Conductance distribution of disordered quasi \\
one-dimensional wires}

%%%%%%%%%%%%%%%%%%%%%%%%%%%%%%%%%%%%%%%%%%%%%%%%%%%%%%%%%%%%%%%%%%%%%%%%%%%%%%
\author{P.\ W\"olfle$^{1,3}$,  and K.\ A.\ Muttalib$^{2,3}$} 
%%%%%%%%%%%%%%%%%%%%%%%%%%%%%%%%%%%%%%%%%%%%%%%%%%%%%%%%%%%%%%%%%%%%%%%%%%%%%%
\newcommand{\address}
  {$^{1}$Institut f\"ur Theorie der Kondensierten Materie, 
  Universit\"at Karlsruhe, D-76128 Karlsruhe, 
  \\ \hspace*{0.5mm} Germany \\ 
  $^{2}$Department of Physics, University of Florida, 
  P.O. Box 118440, Gainesville, FL 32611, USA\\ 
   $^{3}$ Institut f\"ur Nanotechnologie, Forschungszentrum Karlsruhe,
   D-76021 Karlsruhe, Germany}
   
%%%%%%%%%%%%%%%%%%%%%%%%%%%%%%%%%%%%%%%%%%%%%%%%%%%%%%%%%%%%%%%%%%%%%%%%%%%%%%
\newcommand{\email}{\tt woelfle@tkm.physik.uni-karlsruhe.de} 
\maketitle
%%%%%%%%%%%%%%%%%%%%%%%%%%%%%%%%%%%%%%%%%%%%%%%%%%%%%%%%%%%%%%%%%%%%%%%%%%%%%
\begin{abstract}
 We determine analytically the distribution of conductances of quasi
 one-dimensional disordered electron systems, neglecting
 electron-electron interaction, for all strengths of disorder.  We find
 that in the crossover region between the metallic and insulating
 regimes, $P(g)$ is highly asymmetric.  The
 average and the variance of $P(g)$ are shown to agree with exact results. 
\end{abstract}
%%%%%%%%%%%%%%%%%%%%%%%%%%%%%%%%%%%%%%%%%%%%%%%%%%%%%%%%%%%%%%%%%%%%%%%%%%%%%

\section{Introduction}
\label{intro}
The conductance $g$ (in units of $e^2/h$) of a mesoscopic disordered
electron system is known to fluctuate strongly from sample to sample,
or as a function of an external parameter such as a magnetic field or
a gate voltage controlling the electron density [1].  In the metallic
regime these fluctuations are universal and of Gaussian nature,
i.e. the variance of $g$ is given by a pure number independent of the
specifics of the system, depending only on the presence  (or
absence) of time reversal symmetry with respect to orbital or spin
motion (orthogonal, unitary and symplectic cases) [2].  For increasing
disorder the fluctuations grow and are no longer universal, Gaussian
and symmetric about the average value.  When the variance becomes as
large as the average conductance it is necessary to consider the full
distribution of conductances, $P(g)$.  The situation is simple again
in the localized regime, where $P(g)$ is known to be a log-normal
distribution, with variance $\sim < \ell n (1/g)>$ [3].  Except for
numerical studies of finite size systems [4,5,6] little is known
about the conductance distribution in the crossover regime.  These
studies suggest that the distribution is highly asymmetric [5],
with $-\ell nP(g)$ increasing like a power of $g$ for $g\rightarrow
\infty$ and like $(\ell ng)^2$ for $g\rightarrow 0$ [6]. The shape of
$P(g)$ in the crossover regime depends on the spatial dimension,
but appears to be compatible with one-parameter scaling, and hence
universality at a true metal-insulator transition in $d\ge 3$.

On the other hand, analytical results for finite systems (length $L$)
in $d = 2+\epsilon$
dimensions ($\epsilon \ll 1)$, where a weak
disorder approximation can be applied, showed
that the higher moments
of $P(g)$ are non-universal and diverge in the limit $L\rightarrow
\infty$ [7].  It has been proposed, however, that these results are
not incompatible with a universal distribution at the critical point,
which was determined to be a Gaussian with power law tails [8].  This
may seem surprising in view of the numerical results [4,6].  One
should keep in mind, however, that in $d = 2+\epsilon$ dimensions the
critical conductance at the transition is large $<g>_c = 1/\epsilon
\gg 1$, which is deep in the metallic regime and hence is quite
different from the critical value $<g>_c \sim 1$ expected, e.g. in
$d=3$ dimensions.
 
Here we consider the conductance distribution for the simpler case of
a quasi one-dimensional wire of width $W \ll \ell$, where $\ell$ is
the mean free path due to elastic scattering, and length $L\gg \ell$.
Although in this case (for orthogonal and unitary symmetry) all states
are localized in the thermodynamic limit $L \rightarrow \infty$, for
finite length $L$ the system exhibits well defined metallic and
insulating regimes, and a smooth crossover between them.  To be more
precise, this is the case for a quantum wire with ideal leads of the
same cross section, for which the perpendicular momenta at given
energy $E_F$ are quantized into $N$ discrete levels, providing $N$
channels of transport.  The localization length $\xi = N\ell$ in this
case, which for $N \gg 1$ allows for a metallic  regime to be realized
in short wires $(\xi \ll L)$, whereas for long wires $(L \gg \xi)$ the
system is of insulating character.  For strictly one-dimensional
weakly disordered systems the conductance distribution may be obtained
analytically [9], but in this case a metallic regime is absent.  The
dimensionless conductance $g$ of a quantum wire can be expressed in
terms of the N transmission eigenvalues $T_i$ of the corresponding
scattering problem as $g = \Sigma_{i=1}^N T_i$ [10].  The joint
probability distribution $P_T(\{T_i\})$ of the $T_i$ may be obtained
[11] from a Fokker-Planck equation known as the
Dorokhov-Mello-Pereyra-Kumar (DMPK) equation [12], in the limit of
large $N$.  The distribution $P_T(\{T_i\})$ depends only on the
parameter $L/\xi$. The DMPK approach has been shown to be in agreement
with the exact formulation of the problem in terms of a supersymmetric
nonlinear sigma model [13].  Within the latter formulation the average
and the variance of the conductance have been calculated for all
values of $L/\xi$ [14].

To calculate the conductance distribution
$P(g)$ from the joint distribution $P_T(\{T_i\})$ an N-fold
integration is required, subject to the constraints $0 \le T_i \le 1$
and $\Sigma_i T_i = g$, which has only been done in the limiting cases
of $L/\xi \ll 1$ (metal) and $L/\xi \gg 1$ (insulator). Here we describe
a systematic and simple method, valid for all values of $L/\xi$, to
obtain $P(g)$ from $P_T(\{T_i\})$ essentially analytically.  We employ
a generalized saddlepoint approximation, which recovers all the known
results in the limiting cases, and provides results in the crossover
regime in semiquantitative agreement with numerical data and with
analytical results for the average and the variance of $g$.  In
particular, we find that $P(g)$ for $L/\xi \sim 1$ is given by a
``one-sided'' log-normal distribution for $g <1$, with a Gaussian tail
at $g > 1$ [15].
\nopagebreak
\section{Generalized saddlepoint approximation}
\label{saddle}

It is useful to introduce variables $\lambda_i$ and $x_i$ defined by
$T_i = (1 + \lambda_i)^{-1}$, $\lambda_i = \sinh^2 x_i$, in terms of
which the conductance distribution may be represented as
\newpage
\begin{equation}
P(g) = \frac{1}{Z}\int_{-\infty}^\infty \frac{d\tau}{2\pi} e^{i\tau g}
\int_{0}^\infty (\Pi_{i=1}^N d\lambda_i)\exp \Big[-F(\{\lambda_i\};\tau)\Big]
\label{1}
\end{equation}
The ``free energy'' $F$ for unitary symmetry (for orthogonal and
symplectic symmetry the calculation is analogous) is obtained from
the DMPK equation [11]
\begin{equation}
F = 2\sum_i V(\lambda_i) + \sum_{i,j}u(\lambda_i,\lambda_j) + \sum_i
\frac{i\tau}{1 + \lambda_i}
\label{2}
\end{equation}
where $u(\lambda_i,\lambda_j)$ is generated by the Jacobian of the
integration over the transfer matrix elements and leads to ``level
repulsion''. 
Here one may interpret $V(\lambda_i) = (\xi/ 2L) x_i^2$ as a
``one-body potential'' and $u(\lambda_i,\lambda_j) = - \frac{1}{2}
(u_1 +u_2)$, with $u_1(\lambda_i, \lambda_j)= \ell n\mid \lambda_i -
\lambda_j\mid$ and $u_2(\lambda_i, \lambda_j) = \ell n \mid x_i^2 -
x_j^2\mid$ as an ``interaction potential'' of charges at positions
$\lambda_i$.  In the metallic regime $V(\lambda)$ gives rise to a
confinement of the charges in the regime $\lambda_i < 1$ (note
$V(\lambda) \propto \lambda^2$, $\lambda < 1$), such that a
description in terms of a charge density $\rho(\lambda)$ is
appropriate.  In the insulating regime $(V (\lambda) \sim \ell
n^2\lambda, \lambda \gg 1)$ the logarithmic repulsion between the
charges dominates the potential $V(\lambda)$, leading to an
exponentially large separation between the charges, of which only the
one closest to the origin is of importance.\\
\indent To capture both aspects we keep the first eigenvalue $\lambda_1$
separate and represent all the other eigenvalues by a continuum
density
$\rho(\lambda)$, beginning at a lower limit $\lambda_2 > \lambda_1$.
The free energy then takes the form 
\begin{eqnarray}
F(\rho (\lambda); \lambda_1, \lambda_2;\tau)
&=2\int_{\lambda_2}^\infty
d\lambda \rho (\lambda)V_{tot}(\lambda) + 2V(\lambda_1) +\nonumber \\
& + \int_{\lambda_2}^\infty d\lambda d\lambda{'}
\rho(\lambda)u(\lambda,
\lambda{'})\rho(\lambda{'}) + \frac{i\tau}{1 + \lambda_1}
\label{3}
\end{eqnarray} 
where $V_{tot}(\lambda) = V(\lambda) + u(\lambda,\lambda_1) +
\frac{i\tau}{1 + \lambda}$.  The integration over variables
$\lambda_3, ....\lambda_N$ in (1) is replaced by a functional
integration $D[\rho(\lambda)]$.  The latter is done in saddlepoint
approximation, leading to the integral equation for $\rho(\lambda)$
\begin{equation}
\int_0^\infty d\zeta{'} [u_1(\zeta - \zeta{'}) + u_2(\zeta + \lambda_2,
\zeta{'}+\lambda_2)]\rho (\zeta{'} + \lambda_2) = 2V_{tot}(\zeta +
\lambda_2)
\label{4}
\end{equation}
where $\zeta =\lambda - \lambda_2$ has been introduced.  This
integral equation can be solved approximately by putting $u_2(\zeta +
\lambda_2, \zeta{'} + \lambda_2) = u_2(\zeta,\zeta{'}) + \Delta u$ and
neglecting $\Delta u$ in lowest order, which is exact in the limits
$\zeta, \zeta{'} \gg \lambda_2$ and $\zeta, \zeta{'} \ll \lambda_2$.
The leading correction term $\Delta u \propto \lambda_2$ in the
metallic
regime can be treated perturbatively by replacing $\rho$ in the
integral involving $\Delta u$ by the saddlepoint solution for $\Delta
u = 0$, (the results of this approximation will be presented below).\\
\indent
The saddlepoint density $\rho_{sp}(\lambda)$ is found to develop
negative parts for small $\lambda_2$, although $\rho(\lambda)$ is
positive by definition.  We take this as a signal that configurations
of charges with $\lambda_2 < \lambda_c$ (for which $\rho_{sp}
(\lambda)$ starts to turn negative at small $\lambda$) are unphysical
and should be deleted.  This is done by limiting the integration on
$\lambda_2$ to $\lambda_2 > \lambda_c + \lambda_1$.  The free energy
after the saddlepoint integration on $\rho(\lambda)$ is found as 
\begin{equation}
F(\lambda_1,\lambda_2;\tau) = \int_{\lambda_2}^\infty d\lambda V_{eff}
(\lambda)\rho_{sp}(\lambda) + 2V(\lambda_1) + \frac{i\tau}{1 +
\lambda_1} + F_{f\ell}
\label{5}
\end{equation}
where $F_{f\ell}$ is the fluctuation part of the functional integral
on $\rho(\lambda)$ and may be shown to depend on $\lambda_1,
\lambda_2$ as $F_{f\ell} = \frac{1}{2} \ell n (\lambda_2 - \lambda_1)
+ \rm{const}$.

Since $V_{eff}$ and $\rho_{sp}$ are linear functions in $\tau$,
$F(\lambda_1,\lambda_2;\tau) = F^0 + i\tau F{'} +
\frac{1}{2}(i\tau)^2F{''}$ is a quadratic form in $\tau$ leading to a
Gaussian integral over $\tau$ in (1), with the result
\begin{equation}
P(g) = \frac{1}{Z} \int_0^\infty d\lambda_1 \int_{\lambda_1 +
\lambda_c}^\infty d\lambda_2 e^{-S}
\label{6}
\end{equation}
where $S = - (g-F{'})^2/2F{''} + F^0$.  The remaining integrals on
$\lambda_1, \lambda_2$ can be done numerically, or again in saddlepoint
approximation.

\section{Results}
\label{results}

In the metallic regime $(L/\xi \ll 1)$, the relevant values of
$\lambda_1$ and $\lambda_2$ are small of order $L/\xi$.  In the limit
$\lambda_1, \lambda_2 \rightarrow 0$ we find $\mid F{''}\mid =
\frac{1}{15}$ and $F{'} = \xi/L$, whereas $F^0$ tends to a
constant at the saddlepoint $\lambda_2 = \lambda_c + \lambda_1$.  Thus
$P(g)$ is given by a Gaussian centered at $g = \xi/L$, of variance
$1/15$, in agreement with known results [1].\\
\indent We have calculated the expressions for $F^0$ and $F{'}$ analytically
up to and including all terms of order $L/\xi$ (the correction to
$F{''}$ is $0[(L/\xi)^2]$).  The correction to the average conductance
in the metallic regime to this order is found as $< g > = \xi /L -
\eta L/\xi$, with $\eta = 0.027$, which compares well with the exact
result [15] $\eta = 1/45 \simeq 0.022$. There is no correction to
the variance in order $L/\xi$, in agreement with [15].\\
\indent In the insulating regime, $L/\xi \gg 1$, the typical values of $x_1$
and $x_2$ are both $\gg 1$.  In fact, the requirement of positivity of
the density for $L/\xi >, \pi^2/2$ can only be satisfied if $x_2 \rightarrow
\infty$, independent of $x_1$. Using  the saddlepoint values of $x_1 = -
\frac{1}{2}\ell n(g/4)$ and  $F^0 = (\xi/L) x_1^2 - x_1$, $F{'} =
4e^{-2x_1}$ and $F{''} = - \frac{4}{3}e^{-4x_2}\rightarrow 0$ one
finds
\begin{equation}
P(g) = \frac{1}{Z}\frac{1}{g}\exp - [\frac{\xi}{4L} (\ell n(g/4) +
L/\xi)^2]
\label{}
\end{equation}
 a log-normal distribution in agreement with [3].\\  
\indent In the crossover regime on the insulating side, where $\xi/L < 1$, we
make use of the fact that the typical values of $x_1, x_2$ are $x_2
\gg 1$, but $x_1 < x_2$, otherwise arbitrary.  We then find 
%\begin{equation}
$F^0 \simeq (1/3)(\xi/L)^2x_2^3 - (\xi/L)(x_2^2 - x_1^2) + x_2 -
(1/2) \ell n (x_1 \sinh (2x_1))$; 
%\label{8}
%\end{equation}
$F{'}\simeq \cosh^{-2}x_1$ and $F{''} = - \sinh^{-2}(2x_2)[\frac{1}{3}
- \frac{1}{4x_2^2} + \sinh^{-2}(2x_2)]$.

The saddlepoint equation for
$x_1$ is given by $\cosh x_1 = g^{-1/2}$, which has a solution only
for $g \leq 1$.  For $g >1$, instead the boundary values $x_1 = 0$,
$x_2 = (2L/\pi\xi)$ give the minimum of $F$.  The corresponding
results for
$P(g)$ are
\begin{equation}
P(g) = \frac{1}{Z} \exp - a (g-1)^2\;\;\;,\;\;\;g>1
\label{9}
\end{equation}
\begin{figure}[h]
\begin{center}
\includegraphics[width=7cm]{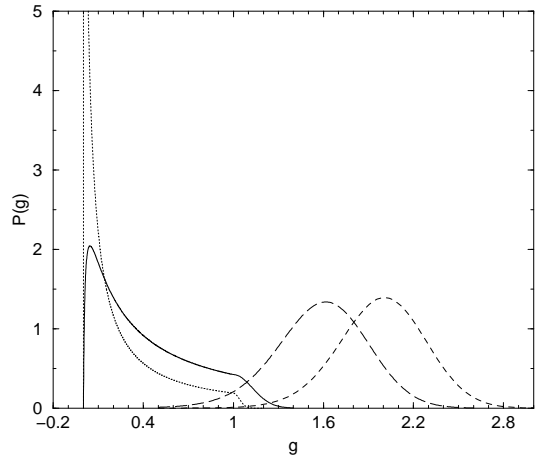}
\end{center}
\caption{Conductance distribution $P(g)$ versus $g$ for $\xi/L = 0.4,
0.7, 1.6, 2.0$ (dotted, solid, long-dashed, short-dashed lines)}
\label{fig1}
\end{figure}
\begin{equation}
P(g) =
\frac{1}{Z}\frac{1}{g}\Big[\frac{\rm{arsech}\sqrt{g}}{g\sqrt{1-g}}
\Big]^{1/2}\exp\Big[-
\frac{\xi}{L}\Big(\rm{arsech}\sqrt{g}\Big)^2\Big],\;\;\; g<1
\label{10}
\end{equation}
Here $a=F{''}(x_2 = 2L/\pi \xi)$ controls the Gaussian cut-off of
$P(g)$
for $g>1$. For $L/\xi \gg 1$ and $g^{-1} \gg 1$ Eq. (9)  reduces 
to the log-normal distribution (7).

\begin{figure}[h]
\begin{center}
\includegraphics[width=7cm]{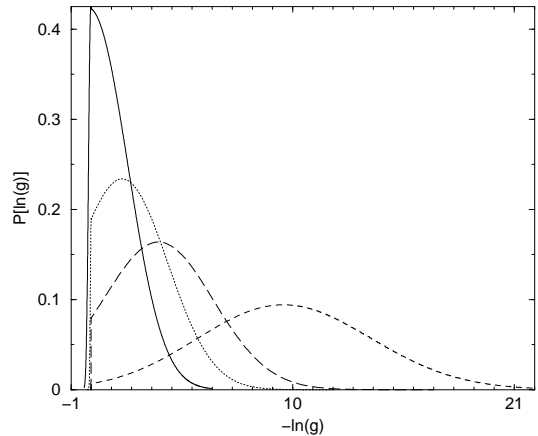}
\end{center}
\caption{Conductance distribution versus $\ell n(1/g)$ for $\xi/L =
0.7, 0.4, 0.25, 0.1$ (solid, dotted, long-dashed, short-dashed lines)}
\label{fig2}
\end{figure}
Fig. 1 shows $P(g)$ versus $g$ for several values of $\xi/L = 0.4,
  0.7, 1.6, 2.0$. In Fig. 2 the results for $\xi/L = 0.4$ and $0.7$
are again shown plotted versus $\ell n(1/g)$ together with results for
$\xi/L = 0.25, 0.1$.
In the logarithmic plot (Fig. 2) one clearly
recognizes a log-normal distribution centered at $\ell n(1/g) \equiv
L/\xi - \ell n 4$ and of variance $var[\ell n(1/g)]\cong 2L/\xi$,
cut-off at $\ell n g = 0 (g=1)$.  The abrupt qualitative change of the
shape of the distribution $P(g)$ in the crossover regime $(L/\xi \sim
1)$ as one goes from values $g < 1$ to $g > 1$ is a consequence of the
small value of $F{''}$.  Note that $\mid F{''}\mid \ll 1$ even in the
metallic regime, decreasing exponentially in the insulating regime.
The term $\propto (g-F{'})^2$ is thus multiplied by the large number
$1/\mid F{''}\mid$, which forces the saddlepoint equation $F{'} = g$,
which, however, has a solution only for $g < 1$.  This leads to a
power law dependence of $S$ on $\ell n g$.  At $g > 1$ the minimum of
$S$ is attained at the boundary of the integration regime $x_1 = 0$,
where the term $\propto (g-F{'})^2$ dominates, resulting in a Gaussian
cut-off of $P(g)$.\\ 
\indent
This result might suggest that quite generally the statistics of the
conductance in the crossover regime is Gaussian centered at $g = 1$
for $g>1$ and log-normal centered at $<\ell n g> = L/\xi$ for $g<1$.
This may be made plausible in the following way.  If the center of the
distribution is located at $g\sim 1$, the lowest eigenvalue $\lambda_1$
still must be dominant, meaning that $\lambda_2 \gg 1$.  The
statistics of $\lambda_1$ is then essentially determined by the single
particle potential $V(\lambda_1)$, giving rise to Gaussian statistics
for $\lambda_1 < 1$, where $V(\lambda_1) \propto \lambda_1^2$ and to
log-normal statistics for $\lambda_1 > 1$, where $V(\lambda_1) \propto
\ell n^2\lambda_1$.  These dependences carry over to the statistics of 
$g \simeq 1/(1+\lambda_1)$.\\
\indent
The shape of $P(g)$ in higher dimensions, as determined numerically
[5,6]
shares the feature of an abrupt change at $g = 1$ from approximately
log-normal to exponential behavior with our results.  Thus, although
the DMPK approach followed here is applicable only for quasi
one-dimensional systems, the qualitative behavior found here may be
more generally valid.
 
\vspace*{0.25cm} \baselineskip=10pt{\small \noindent  
We acknowledge useful discussions with A.D. Mirlin and D.G. Polyakov.
This work has been supported in part by SFB 195 der Deutschen
Forschungsgemeinschaft.}

%%%%%%%%%%%%%%%%%%%%%%%%%%%%%%%%%%%%%%%%%%%%%%%%%%%%%%%%%%%%%%%%%%%%%%%%%%

\end{document}